\documentstyle[12pt]{article}

\newcommand \beq{\begin{eqnarray}}
\newcommand \eeq{\end{eqnarray}}
\begin{document}

\def\grad{\nabla}                               
\def\del{\partial}                              
\def\bfgrad{\mbox{\boldmath$\grad$}}

\def\frac#1#2{{#1 \over #2}}
\def\bra#1{\langle#1\vert}              
\def\ket#1{\vert#1\rangle}              

\def\simge{\mathrel{%
   \rlap{\raise 0.511ex \hbox{$>$}}{\lower 0.511ex \hbox{$\sim$}}}}
\def\simle{\mathrel{
   \rlap{\raise 0.511ex \hbox{$<$}}{\lower 0.511ex \hbox{$\sim$}}}}

%
\def\journal#1#2#3#4{\ {#1}{\bf #2} ({#3})\  {#4}}

\def\AdvPhys{\journal{Adv.\ Phys.}}
\def\AnnPhys{\journal{Ann.\ Phys.}}
\def\EurophysLett{\journal{Europhys.\ Lett.}}
\def\JApplPhys{\journal{J.\ Appl.\ Phys.}}
\def\JMathPhys{\journal{J.\ Math.\ Phys.}}
\def\LettNuovoCimento{\journal{Lett.\ Nuovo Cimento}}
\def\Nature{\journal{Nature}}
\def\NPA{\journal{Nucl.\ Phys.\ {\bf A}}}
\def\NPB{\journal{Nucl.\ Phys.\ {\bf B}}}
\def\NuovoCimento{\journal{Nuovo Cimento}}
\def\Physica{\journal{Physica}}
\def\PLA{\journal{Phys.\ Lett.\ {\bf A}}}
\def\PLB{\journal{Phys.\ Lett.\ {\bf B}}}
\def\PhysRev{\journal{Phys.\ Rev.}}
\def\PR{\journal{Phys.\ Rev.}}
\def\PRC{\journal{Phys.\ Rev.\ {\bf C}}}
\def\PRD{\journal{Phys.\ Rev.\ {\bf D}}}
\def\PRB{\journal{Phys.\ Rev.\ {\bf B}}}
\def\PRL{\journal{Phys.\ Rev.\ Lett.}}
\def\PhysRept{\journal{Phys.\ Repts.}}
\def\ProcNatlAcadSci{\journal{Proc.\ Natl.\ Acad.\ Sci.}}
\def\ProcRoySoc{\journal{Proc.\ Roy.\ Soc.\ London Ser.\ A}}
\def\RevModPhys{\journal{Rev.\ Mod.\ Phys.}}
\def\Science{\journal{Science}}
\def\SovPhysJETP{\journal{Sov.\ Phys.\ JETP}}
\def\SovPhysJETPLett{\journal{Sov.\ Phys.\ JETP Lett.}}
\def\SovJNuclPhys{\journal{Sov.\ J.\ Nucl.\ Phys.}}
\def\SovPhysDoklady{\journal{Sov.\ Phys.\ Doklady}}
\def\ZPhys{\journal{Z.\ Phys.}}
\def\ZPhysA{\journal{Z.\ Phys.\ A}}
\def\ZPhysB{\journal{Z.\ Phys.\ B}}
\def\ZPhysC{\journal{Z.\ Phys.\ C}}

\begin{titlepage}
\begin{flushright} {Saclay-T97/123}
\end{flushright}
\vspace*{2cm}
\begin{center}
\baselineskip=20pt
{\Large{\bf Effective theory for real-time dynamics\\
 in hot gauge theories}}
\vskip0.5cm
Edmond Iancu\\
{\it Service de Physique Th\'eorique\footnote{Laboratoire de la Direction
des
Sciences de la Mati\`ere du Commissariat \`a l'Energie
Atomique}, CE-Saclay \\ 91191 Gif-sur-Yvette, France}\\

\end{center}

\vskip 1cm
\begin{abstract} 

For a high temperature non-Abelian plasma, we reformulate
the hard thermal loop approximation as an effective classical thermal
field theory for the soft modes. 
The effective theory is written in local Hamiltonian form, and the
thermal partition function is explicitly constructed.
It involves an ultraviolet cutoff which separates between
hard and soft degrees of freedom in a gauge-invariant way,
together with counterterms which cancel the cutoff
dependence in the soft correlation functions.
The effective theory is well suited for numerical
studies of the non-perturbative dynamics in real time,
in particular, for the computation
of the baryon number violation rate at high temperature.

 \end{abstract}

\end{titlepage}

\baselineskip=20pt

The violation of the baryon number in the high-temperature, symmetric 
phase of the electroweak theory is an important example of a physical process
which is sensitive to the non-perturbative real-time dynamics of hot
gauge theories \cite{Shapo}. Unlike static characteristics, like the free
energy, which can be computed on the lattice, the non-perturbative evolution 
in real-time cannot be studied through the standard lattice 
simulations formulated in imaginary-time. However, it has long been 
recognized that a fully 
quantum calculation is actually not necessary \cite{Shapo}:  the
non-perturbative phenomena are associated with long wavelength\footnote{$T$ denotes 
the temperature, assumed to be large enough for the
coupling constant $g(T)$ to be small: $g\ll 1$.}
($\lambda \simge 1/g^2T$) magnetic fields, which, because of Bose
enhancement, have large occupation numbers,
\beq\label{BE}
N_0(E)\,\equiv\,\frac{1}{{\rm e}^{\beta E}-1}\,\simeq\,\frac{T}{E}
\qquad{\rm for}\,\,\,\, E\ll T,\eeq
and should therefore exhibit a {classical} behaviour.
Based on this observation, there has been attempts to compute
the baryon number violation rate $\Gamma$ through lattice
simulations of the {\it classical} thermal Yang-Mills theory \cite{AmbK,Turok}.

However, it has been recently observed \cite{Yaffe,Son}
that the baryon violating processes are actually sensitive to the
{\it hard} thermal modes with momenta $\sim T$, for which
the classical approximation is well-known to fail:
the hard modes cause the damping of the soft field
configurations, an effect which is predicted to reduce $\Gamma$
by a factor of $g^2$ as compared with its classical estimate 
in Refs. \cite{Shapo,AmbK}. In order to verify this prediction
and eventually compute $\Gamma$, one has to properly take into account
the effects of the hard modes on the dynamics of the soft fields.
To leading order in $g$, these effects are encompassed by the
so-called ``hard thermal loops'' (HTL) [6--11], which are non-local
one-loop corrections to the soft ($k\simle gT \ll T$) field
propagator and vertices due to the hard 
($k \sim T$) thermal modes.

It has been first suggested in Ref. \cite{McLerran} to use the
local version of the HTL effective theory [8--11]
in classical lattice simulations in order to compute the 
baryon number violation rate.
However, in order to transpose this idea into practice, one first
needs a precise, 
and gauge-invariant, separation between hard and soft degrees of freedom
(to avoid overcounting, and to provide an ultraviolet cutoff
to the effective theory for the soft modes).
Loosely speaking, this requires an intermediate scale $\mu$, 
with $gT \ll \mu \ll T$, which should act as an infrared (IR) 
cutoff for the hard modes and as an ultraviolet (UV) cutoff
for the soft ones,
and which should cancel in the calculation of physical quantities.
But the practical implementation of such a separation of scales meets with
technical difficulties:
(i) In the hard sector, one cannot simply introduce $\mu$ as an
infrared cutoff in the one-loop diagrams for the HTL's since this would
break gauge symmetry \cite{McLerran}. (ii) In the soft sector, one cannot
use the lattice spacing to provide the necessary upper cutoff $\sim \mu$:
indeed, a finite (and relatively large: $a \sim \mu^{-1}
\gg T^{-1}$) lattice spacing introduces lattice artifacts
which make impossible the matching with the hard sector
\cite{McLerran,Arnold}.

Another problem is the proper definition of classical
thermal expectation values within the effective theory.
In order to compute such expectation values,
one has to solve first the equations of motion for given 
initial conditions, and then average over the classical phase
space with the Boltzmann weight exp$(-\beta H)$. Since
the local formulations of the HTL theory which are currently
available \cite{qcd,Nair,emt} are not in canonical form, 
it is a non-trivial task to identify the independent degrees of freedom
and construct the classical phase space.

The purpose of this Letter is to give an explicit solution to the above
mentioned problems by constructing a local effective theory for the
soft modes with a $\mu$-dependent Hamiltonian.
The classical partition function will involve an explicit UV cutoff, 
chosen so as to cancel the $\mu$-dependence of the Hamiltonian in the 
calculation of IR-sensitive correlation functions, like the
baryon number violation rate $\Gamma$.

Our construction relies in an essential way on the local formulation
of the HTL theory presented in Ref. \cite{qcd},
which we briefly review now.
It involves a set of coupled equations
for the soft fields and their induced current, namely
eqs.~(\ref{ava})--(\ref{N}) below. The
 soft fields $A_\mu^a(x)$ satisfy the Yang-Mills
equations with an induced current in the right hand side:
\beq\label{ava}
(D_\nu F^{\nu\mu})_a\,=\,j^\mu_a,
\eeq
where $D_\mu = \del_\mu+ig[A_\mu,\,\cdot\,]$, $A_\mu = A_\mu^a T^a$, and
$F_{\mu\nu}= [D_\mu, D_\nu]/(ig)$. (The generators of the colour group
in the adjoint representation are denoted by $T^a$; they satisfy
$[T^a,T^b]=i f^{abc}T^c$ and
Tr$(T^aT^b)=C_A\delta^{ab}$, with $C_A=N$ for SU($N$).)
The induced current $j_\mu= j_\mu^a T^a$ 
is  related to the colour fluctuations of the hard thermal modes:
\beq\label{j}
j^{\mu}_{a}(x)
=2gC_A\int\frac{{\rm d}^3k}{(2\pi)^3}\,v^\mu
\,\delta N_a({\bf k},x).\eeq
In this equation, $\delta N({\bf k},x)= \delta N_a({\bf k},x)\,T^a$
is a phase-space colour density matrix for hard gluons ($|{\bf k}|\sim T$),
which describes long-wavelength colour correlations as
induced by the soft fields $A_\mu^a$.
 Furthermore, $v^\mu= (1,\,{\bf v})$ and ${\bf v}=
{\bf k}/k$ is the velocity of the hard particle ($k= |{\bf k}|$,
and $|{\bf v}| =1$). The system is closed by the kinetic equation
for the density matrix, which is a non-Abelian generalization of
the Vlasov equation:
\beq\label{N}
(v\cdot D_x)\delta N({{\bf k}},x)\,=\,-\, g\,
{\bf v}\cdot{\bf E}(x)\,\frac{{\rm d}N_0}{{\rm d}k}\,,\eeq
where $E_a^i\equiv F_a^{i0}$ and $N_0(k)= 1/({\rm e}^{\beta k}-1)$.
The dynamics described by the above equations is gauge invariant, and
the current $j^a_\mu$ is covariantly conserved: $D^\mu j_\mu =0$.

From eq.~(\ref{N}), we note that the ${\bf v}$ and $k$-dependence 
can be factorized in $\delta N^a({{\bf k}},x)$ by writing:
\beq\label{dn}
\delta N^a({\bf k}, x)\equiv - gW^a(x,{\bf v})\,({\rm d}N_0/{\rm d}k).\eeq
The new functions $W^a(x,{\bf v})$ satisfy the equation:
\beq\label{W}
(v\cdot D_x)W(x,{\bf v})\,=\,{\bf v}\cdot{\bf E}(x),\eeq
which is independent of $k$ since the hard particles move at the speed
of light: $|{\bf v}| =1$. 

Eqs.~(\ref{ava})--(\ref{W}) above
provide a local description of the soft field dynamics in the
HTL approximation. In order to use these equations for classical
thermal calculations, one needs to (i) introduce an IR cutoff $\mu$
in the hard sector, (ii) perform a Hamiltonian analysis (to identify
the independent degrees of freedom and the corresponding Hamiltonian), 
(iii) write down the classical partition function, and (iv)
supply the effective theory with an UV cutoff $\sim \mu$.
 We shall address these problems in this order:

(i) By inspection of eqs.~(\ref{ava})--(\ref{N}), it is quite
 obvious how to introduce the intermediate scale $\mu$: 
since ${\bf k}$ is the momentum carried by the hard particles,
it is sufficient to integrate in  eq.~(\ref{j}) with a lower cutoff
equal to $\mu$. With eq.~(\ref{dn}), the radial integration in eq.~(\ref{j})
can be worked out, with the result
\beq\label{j1}
j^\mu_a(x)&=&m_H^2(\mu)\int\frac{{\rm d}\Omega}{4\pi}
\,v^\mu\,W_a(x,{\bf v}),\eeq
where the angular integral $\int {\rm d}\Omega$ runs over the unit sphere 
spanned by ${\bf v}$, and
\beq\label{MDMU}
m^2_H(\mu)\equiv -\frac{g^2 C_A}{\pi^2}\int_{\mu}^\infty 
{\rm d}k k^2\frac{{\rm d}N_0}{{\rm d}k}\simeq\frac{g^2 C_A}{3}\left(
T^2-\frac{3}{\pi^2}\mu T\right).\eeq
The quantity $m_D\equiv m_H(\mu=0)$ is the physical Debye mass
to leading order in $g$  \cite{BP90,FT90,qcd}.
For $gT \ll \mu \ll T$, $m_H(\mu)$ is the hard sector contribution to $m_D$.

In contrast to the usual one-loop  calculations \cite{McLerran}, the
above implementation of $\mu$ has preserved gauge symmetry automatically:
indeed, the kinetic equation (\ref{N}) is gauge covariant for any value of $k$,
so that the $\mu$-dependent current in eq.~(\ref{j1}) is covariantly conserved.

(ii) A Hamiltonian analysis of the HTL theory has been given 
by Nair \cite{Nair}, in terms of some new auxiliary fields.
Here, we shall rather follow Refs. \cite{emt} and propose a simpler
Hamiltonian formulation which involves the fields $W_a(x,{\bf v})$
introduced above. In the gauge $A^a_0=0$,
the independent degrees of freedom are 
$E^a_i$, $A^a_i$ and  $W^a$, and the corresponding equations of motion 
follow from eqs.~(\ref{ava}), (\ref{W}) and (\ref{j1}) above:
\beq\label{CAN}
E^a_i&=&-\del_0 A^a_i,\nonumber\\
-\del_0 E^a_i +\epsilon_{ijk}(D_j B_k)^a &=&
m_H^2(\mu)\int\frac{{\rm d}\Omega}{4\pi}\,v_i\,W^a(x,{\bf v}),\nonumber\\
\left(\del_0 + {\bf v\cdot D}\right)^{ab} W_b&=&{\bf v \cdot E}^a,\eeq
together with Gauss' law which in this gauge must be imposed as a constraint:
\beq\label{GAUSS}
({\bf D\cdot E})^a\,+\,m_H^2(\mu)\int\frac{{\rm d}\Omega}{4\pi}\,W^a(x,{\bf v})
\,=\,0.
\eeq
Note that eqs.~(\ref{CAN}) are not in canonical form: this is already
obvious from the fact that we have an odd number of equations.
Accordingly, it is not a priori clear that these equations are
Hamiltonian in any sense\footnote{Recall that the Hamiltonian structure
is a non-trivial issue already for the Abelian Maxwell-Vlasov equations
\cite{MR}, of which eqs.~(\ref{CAN}) can be seen as a non-Abelian
generalization.}. It has been shown in Ref. \cite{emt} that eqs.~(\ref{CAN})
are conservative: the associated, conserved energy can be computed
in any gauge as:
\beq\label{H}
H\,=\,\frac{1}{2}\int {\rm d}^3 x\left\{{\bf E}_a\cdot{\bf E}_a\,+\,
{\bf B}_a\cdot{\bf B}_a\,+\,m_H^2(\mu)
\int\frac{{\rm d}\Omega}{4\pi}\,W_a(x, {\bf v})\,W_a(x, {\bf v})\right\}.\eeq
Remarkably, we show now 
that, in the gauge $A_0^a=0$, the functional (\ref{H}) also acts
as a Hamiltonian, that is, as a generator of the time evolution.
To this aim, we introduce the following Poisson brackets\footnote{More
precisely, these are generalized Lie-Poisson brackets, according to the
terminology in Ref. \cite{MR}.} (see also Ref. \cite{Nair}) :
\beq\label{PB}
\left\{E^a_i({\bf x}), A^b_j({\bf y})\right\}&=&-\,\delta^{ab}\delta_{ij}
\delta^{(3)}({\bf x-y})\,,\nonumber\\
\left\{E^a_i({\bf x}), W^b({\bf y,v})\right\}&=&v_i\,\delta^{ab}
\delta^{(3)}({\bf x-y})\,,\nonumber\\
m^2_H\left\{W^a({\bf x,v}), W^b({\bf y,v'})\right\}&=&
\left(gf^{abc}W^c+({\bf v\cdot D}_x)^{ab}\right)\delta^{(3)}({\bf x-y})
\delta({\bf v},{\bf v}^\prime)\,.\eeq
Here, $\delta({\bf v},{\bf v}^\prime)$ is the delta function on the unit sphere,
normalized such that
\beq
\int\frac{{\rm d}\Omega}{4\pi}\,\delta({\bf v},{\bf v}^\prime)
\,f({\bf v})\,=\,f({\bf v}^\prime),\eeq
and all the other Poisson brackets are assumed to vanish.
We also assume standard properties for such brackets, namely
antisymmetry, bilinearity and Leibniz identity. It is then
 straightforward to verify that (a) the Poisson
brackets (\ref{PB}) satisfy the Jacobi identity (as necessary for
consistency) and (b) the equations of motion (\ref{CAN}) follow as
canonical equations for the Hamiltonian (\ref{H}). For instance,
$\del_0 W^a=\{H,W^a\}$, and similarly for $E_i^a$ and $A_i^a$.

Note that the effective theory
in eqs.~(\ref{CAN})--(\ref{PB}) involves the infrared cutoff $\mu$
(and also the temperature $T$) only through a single mass parameter,
namely the ``hard'' Debye mass $m_H^2(\mu)$ of eq.~(\ref{MDMU}).

(iii) We are now in position to construct (generally time-dependent)
thermal expectation values within the classical field theory defined by
eqs.~(\ref{CAN})--(\ref{H}). The thermal phase-space is defined
by the initial conditions for eqs.~(\ref{CAN}), and the canonical weight
is given by the effective Hamiltonian (\ref{H}). Thus,
the thermal correlation functions of the fields
$A^i_a$ can be obtained from the following generating functional:
\beq\label{Z}
Z_{cl}[J^a_i]\,=\,
\int {\cal D}{\cal E}^a_i\,{\cal D}{\cal A}^a_i\,{\cal D}{\cal W}^a\,
\delta({\cal G}^a)\,
\exp\left\{-\beta H\,+\,\int{\rm d}^4x J^a_i(x) A^a_i(x)\right\},\eeq
where $A^i_a(x)$ is the solution to eqs.~(\ref{CAN})
with the initial conditions $\{{\cal E}^a_i,{\cal A}^a_i,{\cal W}^a\}$
(that is, $E^a_i(t_0,{\bf x})={\cal E}^a_i({\bf x})$, etc., with 
 arbitrary $t_0$), and $H$
is expressed in terms of the initial fields.
Since the dynamics is gauge-invariant, it is sufficient to enforce Gauss' law
 at $t=t_0$:
\beq {\cal G}^a\,\equiv\,
({\cal D}_i {\cal E}_i)^a\,+\,m_H^2(\mu)\int\frac{{\rm d}\Omega}{4\pi}\,
{\cal W}^a\,=\,0.\eeq

The only subtle point in eq.~(\ref{Z}) is the definition of the measure
in the phase-space\footnote{I am grateful to
Tanmoy Bhattacharya for an illuminating
discussion on this point.}: since we are not using canonical variables,
we still have to verify that the na\"{\i}ve measure
${\cal D}{\cal E}^a_i{\cal D}{\cal A}^a_i{\cal D}{\cal W}^a$ 
is indeed the correct one. A necessary condition is that this measure
be invariant under the time evolution described by eqs.~(\ref{CAN}),
so that $Z_{cl}[J]$ be independent of $t_0$, as it should.
This condition can be most easily verified by considering an infinitesimal
time evolution of the form
$\Phi_\alpha\,\to\,\Phi_\alpha^\prime\equiv\Phi_\alpha + \{H, \Phi_\alpha\}
{\rm d}t$, where $\Phi_\alpha$ refers to any of the field variables
$\{{\cal E}^a_i,{\cal A}^a_i,{\cal W}^a\}$. Then, by using 
eqs.~(\ref{CAN}) --- or, equivalently, the Poisson brackets (\ref{PB})
---, it is straightforward to verify that the Jacobian for this
transformation is equal to one, to linear order in ${\rm d}t$ :
\beq
{\cal J}\,\equiv\,\left|\frac{\delta
({\cal E}^\prime_i,{\cal A}^\prime_i,{\cal W}^\prime)}{\delta
({\cal E}_i,{\cal A}_i,{\cal W})}\right|\,=\,1\,+\,O(({\rm d}t)^2).\eeq

As a further check, one can verify that eq.~(\ref{Z}) reduces to standard
results in some particular cases. For instance, for $J^a_i=0$ this equation
yields the result expected from dimensional reduction \cite{Kajantie}, that is,
\beq\label{ZRED}
Z_{cl}\,=\,\int {\cal D}{\cal A}^a_0\,{\cal D}{\cal A}^a_i\,
\exp\left\{-\frac{\beta }{2}\int{\rm d}^3x \,\Bigl(
{\cal B}^a_i{\cal B}^a_i + ({\cal D}_i {\cal A}_0)^a
({\cal D}_i {\cal A}_0)^a+ m_H^2(\mu){\cal A}_0^a{\cal A}_0^a
 \Bigr)\right\},\eeq
where the ${\cal A}_0^a$ components of the gauge fields have been reintroduced
as Lagrange multipliers to enforce Gauss' law, and the functional
integrals over ${\cal E}^a_i$ and ${\cal W}^a$ have been explicitly performed.
Converserly, the general formula (\ref{Z})
can be seen as a generalization of the dimensional reduction method to
include dynamical (i.e., time-dependent) phenomena.
Another check is provided by the Abelian limit, where eq.~(\ref{Z}) yields,
after a straightforward calculation :
\beq\label{ZAB}
Z_{cl}[J_i]&=&\exp\left\{-\frac{1}{2}\int{\rm d}^4x \int{\rm d}^4y
\, J_i(x) \,{}^*D_{ij}(x-y) J_j(y)\right\},\nonumber\\
{}^*D_{ij}(x-y)&\equiv&\int\frac{{\rm d}^4q}{(2\pi)^4}\,{\rm e}^{-i q\cdot(x-y)}\,
\,{}^*\rho_{ij}(q) N_{cl}(q_0),\eeq
where ${}^*\rho_{ij}(q)$ is the magnetic photon spectral density
in the HTL approximation \cite{BP90} and
$N_{cl}(q_0)\equiv T/q_0$ is the classical thermal distribution function,
which coincides with the low energy limit of the quantum
distribution (cf. eq.~(\ref{BE})).
In the second line of eq.~(\ref{ZAB}) we recognize,  as expected, 
the classical limit of the soft photon 2-point function
in the HTL approximation.

(iv) The last step toward a well-defined classical
effective theory is to supply the partition function (\ref{Z})
with an ultraviolet cutoff  $\Lambda_{cl}\sim \mu$,
chosen so as to cancel --- in the calculation of the soft correlation 
functions --- the explicit $\mu$-dependence of the effective Hamiltonian
(\ref{H}) (a cancellation to be subsequently referred to as {\it matching}). 
Since the effective theory is ultimately intended for non-perturbative
calculations, it will be convenient to chose a regularization
method which can be also implemented on a lattice. This cannot be
the lattice spacing itself: indeed, if we choose to work
with a finite lattice spacing $a\sim 1/\mu$, 
then we break rotational and dilatation symmetry, and the UV 
structure of the lattice theory gets so complicated 
that the matching cannot be performed anymore
\cite{McLerran,Arnold}. Lattice 
artifacts can be eliminated only by taking the continuum limit 
$a\to 0$, which requires $a$ to be independent of $\mu$.

The strategy that we propose here is thus the following:
the effective theory will be formulated as a cutoff theory {\it 
in the continuum}, and the matching will be performed in the continuum,
rotationally-invariant theory. Then, for computational purposes,
the resulting cutoff theory must be put
on a lattice with small lattice spacing $a\ll 1/\Lambda_{cl}$
\footnote{Note that a similar strategy has been recently proposed by Arnold,
in the context of purely classical Yang-Mills theory \cite{Arnold};
of course, the matching was not an issue in Ref. \cite{Arnold},
which was rather concerned with improving the rotational
symmetry in classical lattice simulations.}. 
Because of the explicit UV cutoff  $\Lambda_{cl}$, the continuum
limit $a\to 0$ is well-defined. 

Following Ref. \cite{Arnold},
we introduce a smooth UV cutoff in the continuum theory by replacing,
in the effective Hamiltonian (\ref{H}),
\beq\label{REG}
{\rm Tr}\, B^i B^i\,\longrightarrow\,{\rm Tr}\, B^i f\left(\frac
{{\bf D}^2}{\Lambda_{cl}^2}\right) B^i,\eeq
where $f(z)=1+z^2$ and the trace refers to color indices.
Besides being gauge-invariant, the regularization prescription in 
eq.~(\ref{REG}) has also the advantage that it can be carried out
on the lattice (by using improved lattice Hamiltonians \cite{Arnold,Moore1}). 
However, this prescription breaks down dilatation symmetry and,
as a consequence, the matching cannot be performed for {\it all} the soft
correlation functions (see below), but only for
the non-perturbative quantities which are infrared sensitive,
like the baryon number violation rate $\Gamma$.
Let us explain this in more detail:

For matching purposes,
we need the $\Lambda_{cl}$-dependent corrections to the soft
correlation functions to one-loop order in the effective theory.
Rather than computing loop diagrams, it is more
convenient to rely on kinetic theory to describe the interactions
between the genuinely soft fields, with momenta $k\simle gT$, and the
relatively ``hard'' classical modes, with momenta $k\simge \Lambda_{cl}$.
The relevant kinetic equations can be derived in the same way \cite{qcd},
and look similarly, to our previous equations (\ref{ava})--(\ref{W}).
The only differences refer to the replacement of $N_0(k)$ 
by $N_{cl}(E_k)\equiv T/E_k$,
and of the unit vector ${\bf v}$ by the group velocity
${\bf v}_k\equiv \bfgrad_k E_k$.
(These prescriptions can be readily verified by inspecting
the derivation of the kinetic equations in Ref. \cite{qcd}; they have been
also justified by a diagrammatic analysis in Ref. \cite{Arnold}.)
Here, $E_k^2=k^2 f(k^2/\Lambda^2_{cl})$ is the dispersion equation for
the relatively ``hard'' ($k\simge \Lambda_{cl}$) classical excitations, 
for which  the HTL corrections are relatively small (since
$m_H \sim gT \ll \Lambda_{cl}$) and can be neglected to the order of interest.
It then follows that the colour current due to the ``hard'' classical modes
--- and which summarizes the $\Lambda_{cl}$-dependence of the effective
theory to one loop order --- has the form (compare to 
eqs.~(\ref{j1})--(\ref{MDMU})):
\beq\label{jS}
j^{\mu\,a}_S(x)&=&-2g^2C_A\int\frac{{\rm d}^3k}{(2\pi)^3}
\,\frac{{\rm d}N_{cl}}{{\rm d}E_k}\,v^\mu_k\,W^a(x,{\bf v}_k),\eeq
with the functions $W^a(x,{\bf v}_k)$ satisfying (in the temporal gauge
$A_0=0$):
\beq\label{WS}
\left(\del_0 + {\bf v}_k\cdot{\bf D}\right)^{ab} W_b(x,{\bf v}_k)
&=&{\bf v}_k \cdot{\bf E}^a(x).\eeq
An important difference with respect to eq.~(\ref{j1}) is that
the classical ``hard'' excitations do not move at the speed of light,
but rather with a $k$-dependent velocity ${\bf v}_k$. Accordingly,
the radial and angular integrations in eq.~(\ref{jS}) cannot be 
disentangled anymore, and the current $j^{\mu\,a}_S$ 
will {\it not} be characterized, in general, by a single
mass scale (in contrast to the HTL current in eq.~(\ref{j1})),
which prevents us from performing a full matching.

However, as we show now, the matching can still be done
in the calculation of non-perturbative quantities which are infrared sensitive.
It is indeed well-known (see, e.g., Refs. \cite{Yaffe,Son,Arnold,lifetime})
that the field configurations which are responsible
for the non-perturbative phenomena (and also for the IR divergences of
the perturbation theory) are very soft ($k\equiv |{\bf k}|
\sim g^2 T$) magnetic fields of almost zero frequency: $k_0\simle
g^4 T \ll k$. Indeed, these are the only configurations which are not
screened at the scale $gT$ by the HTL's. For such fields, time derivatives
are suppressed with respect to spatial gradients, and eq.~(\ref{WS})
reduces to:
\beq\label{WS1}
({\bf v}\cdot{\bf D}) W(x,{\bf v}_k) &=&{\bf v} \cdot{\bf E}(x),\eeq
where we have been able to simplify one factor of $|{\bf v}_k|$,
so that ${\bf v}$ is an unit vector, as
in eq.~(\ref{W}). Eqs.~(\ref{WS1}) shows that
on the relevant, non-perturbative field configurations,
the function $ W(x,{\bf v}_k)
\equiv  W(x,{\bf v})$ is independent of the radial momentum
$k$. Then, the radial integral in eq.~(\ref{jS}) factorizes
and yields (for the relevant, magnetic piece of the current):
\beq\label{jS1}
j^{i\,a}_S(x)\,\simeq\,m_S^2(\Lambda_{cl})
\int\frac{{\rm d}\Omega}{4\pi}\,v^i\,W^a(x,{\bf v}),\eeq
with (compare to eq.~(\ref{MDMU})):
\beq\label{A}
m_S^2(\Lambda_{cl})&\equiv&-\frac{g^2 C_A}{\pi^2}\int_0^\infty 
{\rm d}k\,k^2\,|{\bf v}_k|\,
\frac{{\rm d}N_{cl}}{{\rm d}E_k}\nonumber\\
&=&\frac{2g^2 C_A}{\pi^2}\int_0^\infty {\rm d}k\,k\,N_{cl}(E_k)
\,=\,{\kappa} g^2 C_A T\Lambda_{cl}\,.\eeq
The precise value of the numerical coefficient $\kappa$ 
can be found in eq.~(4.8) of Ref. \cite{Arnold}.

Eq.~(\ref{jS1}) shows that, to one-loop order in the effective theory,
the $\Lambda_{cl}$-dependent corrections
to the amplitudes involving quasistatic and soft ($k_0\ll k\sim g^2T$)
magnetic fields are characterized by a single mass scale, namely
$m_S^2(\Lambda_{cl})$.
This is similar to the HTL current in eq.~(\ref{j1}),
so it is now possible to perform the matching by requiring
$m_S^2(\Lambda_{cl})$, eq.~(\ref{A}), to cancel the $\mu$-dependent 
piece of $m_H^2(\mu)$, eq.~(\ref{MDMU}). This is achieved
by choosing $\mu=\pi^2\kappa\Lambda_{cl}\,$:
with this matching condition, the IR-sensitive quantities
computed in the effective theory (\ref{CAN})--(\ref{Z})
with the UV regularization (\ref{REG}) come out independent
of $\mu$ and $\Lambda_{cl}$, for $\mu$ in a large range of values:
$gT \ll \mu \ll T$.

In particular, the effective theory thus defined provides
a $\mu$-independent value for the baryon number violation rate $\Gamma$, 
to be ultimately computed on the lattice.
The lattice implementation of the effective theory (which requires
a lattice version of the new fields $W^a(x,{\bf v})$ and of the
corresponding equations of motion) is by itself a non-trivial issue,
which remains beyond the scope of the present work. 

Let us finally note a different proposal \cite{Hu} for including 
the HTL's, which is to treat the hard degrees of freedom 
as classical coloured particles \cite{Liu}. This
has been recently implemented in lattice simulations \cite{Moore},
with results which seem to confirm the predictions in Ref. \cite{Yaffe}.
By comparaison, the method that we have proposed here,
besides being derived from first principles, has also the advantages
to give the hard modes the correct quantum statistics,
to involve no free parameter, and to perform a precise
matching between hard and soft degrees of freedom, thus allowing
for the continuum limit to be taken in lattice simulations.

To conclude, we have provided an effective classical thermal field 
theory for the soft modes which includes the hard modes in the HTL
approximation and which is well-suited for numerical studies
of the real-time non-perturbative dynamics. 
Important applications include the high-$T$ anomalous baryon number violation,
the dynamics of the electroweak phase transition (which requires
adding the Higgs field to the above theory), and non-perturbative
properties of a hot quark-gluon plasma.

\bigskip
{\noindent {\bf Acknowledgements}}:
During the elaboration of this paper, I have benefited
from discussions and useful remarks from a number
of people. It is a pleasure to thank
J. Ambj{\o}rn, T. Bhattacharya, J.P. Blaizot, S. Habib, 
A. Krasnitz, L. McLerran,
E. Mottola, J.Y. Ollitrault, M.E. Shaposhnikov, A. Smilga and N. Turok.

\end{document}